\documentclass[12pt]{amsart}

\usepackage{amsmath}
\usepackage{latexsym}
\usepackage{amsfonts}
\usepackage{amssymb}
\newcommand{\cal}{\mathcal}

\newcommand{\lh}{{\cal L}({\cal H})}

\newcommand{\hi}{{\cal H}}

\newcommand{\ip}[2]{\left\langle\,#1\,|\,#2\,\right\rangle}
\newcommand{\ket}[1]{\mid#1\rangle}
\newcommand{\kb}[2]{|#1\,\rangle\langle\,#2|}
\newcommand{\fii}{\varphi}

\newcommand{\aspp}{A^{\ket s}_{\psi,\psi}}
\newcommand{\asff}{A^{\ket s}_{\fii,\fii}}
\newcommand{\asmn}{A^{\ket s}[m,n]}
\newcommand{\asnn}{A^{\ket s}[n,n]}
\newcommand{\agmn}{A^{\ket 0}[m,n]}
\newcommand{\asnm}{A^{\ket s}[n,m]}
\newcommand{\uqs}{E^{(Q,s)}}
\newcommand{\ups}{E^{(P,s)}}
\newcommand{\uqsf}{E^{(Q,s)}_{\fii,\fii}}
\newcommand{\upsf}{E^{(P,s)}_{\fii,\fii}}
\newcommand{\gqsf}{g^{(Q,s)}_{\fii,\fii}}

\newcommand{\R}{\mathbb R}

\newtheorem{theorem}{Theorem}[section]
\newtheorem{lemma}[theorem]{Lemma}%[section]
\newtheorem{remark}[theorem]{Remark}%[section]
\newtheorem{corollary}[theorem]{Corollary}%[section]

\begin{document}

\title[]{The uniqueness question in the multidimensional moment problem with
applications to phase space observables}

\author{Anatolij  Dvure\v censkij}
\address{Anatolij  Dvure\v censkij,
Mathematical Institute, Slovak Academy of Sciences,
SK--814~73 Bratislava, Slovakia}
\email{dvurecen@mau.savba.sk}
\author{Pekka Lahti}
\address{Pekka Lahti,
Department of Physics, University of Turku, FIN-20014 Turku, Finland}
\email{pekka.lahti@utu.fi}
\author{Kari Ylinen}
\address{Kari Ylinen, Department of Mathematics, University of Turku, FIN-20014 Turku, Finland}
\email{ylinen@utu.fi}

%\date{\today}

\begin{abstract}
The theory of holomorphic functions of several complex variables is applied
in proving a multidimensional variant of a theorem involving 
 an exponential  boundedness  criterion 
%\cite{BergChr} 
for the classical  moment problem. 
A theorem of Petersen
%\cite{Petersen} 
concerning the relation between the multidimensional  and  one-dimensional
moment problems is extended for half-lines and compact subsets of the real line $\mathbb R$.
These results are used to solve the moment problem for the  quantum phase space observables
generated by the number states.

\end{abstract}
\maketitle{}

\section{Introduction and notations}

The need to regard quantum observables
as positive normalized operator measures, as opposed to the more traditional
spectral measure approach,
motivates the study of
 the moment operators of such observables, and
in particular raises
the question of the uniqueness of the observable
given its moment operators.
The spectral theorem for self-adjoint operators
suffices to exhaust these problems in the case of
spectral measures; in particular, the first moment of a spectral
measure already determines it uniquely.

An important class of quantum observables that are not spectral measures
consists of certain  phase space
observables. These
have proved highly useful in several  branches of quantum physics,
including quantum
communication and information theory,
quantum optics and quantum measurement theory.
Especially
the possibility of experimental implementation
of such observables by modern technology
has drawn a lot of attention to their study.

The original motivation for the research reported in this paper came
from the desire to shed light on the problem of the moment operators
of phase space observables. This is intimately connected with the
general multidimensional moment problem, whose study
in our presentation occupies Sections 2 and 3.
The choice of material in this part is basically dictated by
applications to phase space observables, though not all the results
are strictly needed in the sequel.

We denote as usual $||x|| = (x_1^2 +\cdots + x_n^2)^{1/2}$
for  $x=(x_1,\dots,x_n)\in \mathbb R^n$.
If $K$ is a nonempty Borel subset of $\mathbb R^n$,
we let ${\cal B}(K)$ denote
its Borel $\sigma$-algebra
 and let ${\cal M}^*_n(K)$ be the set of all
measures  $\mu:{\cal B}(K)\to[0,\infty)$
satisfying $\int_K ||x||^{2k} d\mu(x) <\infty $ for $k = 0,1,2,\ldots.$
We write $\mathbb N_0 :=\{0,1,2,\ldots\}.$
For $\mu\in {\cal M}^*_n(K)$ and $k= (k_1,\ldots,k_n) \in \mathbb N_0^n$,
we define the {\it moment} $c_k(\mu)$ as follows:
$$ c_k(\mu) = c_{k_1\ldots k_n}(\mu):=  \int_K x^{k} d\mu(x)
=\int_K x_1^{k_1}\cdots x_n^{k_n} d\mu(x_1,\ldots,x_n).
%,\eqno(2.0)
$$
The multidimensional
moment problem is to find
conditions on a (multi)sequence
$(c_k)_{k \in \mathbb N_0^n}$  under which  there exists a measure
$\mu:{\cal B}(K)\to[0,\infty)$ such that $c_k=c_k(\mu)$ for all
$k\in \mathbb N_0^n$.
It is known that a measure $\mu$ need not be
uniquely determined by its moment sequence $(c_k)_{k \in \mathbb N_0^n}.$
For $\mu\in {\cal M}^*_n(K)$, we denote
$$ V[K,\mu] :=\{\nu \in {\cal M}^*_n(K):
c_k(\nu) = c_k(\mu)
%\  \int_K x^{k} d\nu(x) = \int_K x^{k} d\mu(x),
\ \text{ for all }
k \in \mathbb N_0^n\}.
$$
%Then $V[K,\mu] $ is a nonvoid convex set.
We say that $\mu$ is {\it determined on $K$
by its moment sequence} if $V[K,\mu]$ is a singleton.
In this situation we also say that $\mu$ is {\it determinate}
(on $K$).

As usual, for $p\geq1$,
 $L^p(K,\mu)$ will denote the space of
all (equivalence classes of) Borel 
functions $f:\ K \to\mathbb R$ satisfying
$\int_K  |f(x)|^p d\mu(x) < \infty. $ Let ${\cal P}_n$
denote the set of all polynomials in $x_1,\ldots,x_n$ (or
also the set of their restrictions to a subset
of $\mathbb R^n$ clear from the context).

In Section 2 we prove a multidimensional generalization of an exponential boundedness
criterion in the classical moment problem,
a result involving  \cite[Theorem 6]{BergChr}. Section 3 extends the
results of Petersen \cite{Petersen} on the relation between the multidimensional and the
one-dimensional moment problems. Section 4 introduces the phase space observables,
Section 5 investigates their moment operators, and Section 6 shows the uniqueness of the 
number state generated phase space observables in view of their moment sequences.
In the final sections  the same results are obtained using the Cartesian margins (Section 7) and the polar margins
(Section 8) of the phase space observables.

\section{Uniqueness in the multidimensional moment problem:
exponentially bounded measures}

The theorem of this section is a multidimensional
generalization of \cite[Theorem 6]{BergChr}.
Our proof also resembles that of \cite{BergChr}
(which according to the authors is inspired by
\cite{Hew}),
but in our multidimensional case the theory of holomorphic
functions of several complex variables is used.
It is worth noting that even before \cite{Hew},
a closely related proof was given in 1950 in the Russian
original of \cite{AG1,AG2}, see
\cite[p. 25--26]{AG1}. We call {\it exponentially bounded}
the type of measures appearing in the next result.

\begin{theorem}\label{T1}
Let $\mu:\mathcal B(\mathbb R^n)\to [0,\infty)$ be a measure such that
\begin{equation}\label{E1}
\int_{\mathbb R^n} e^{a\parallel x\parallel}\,d\mu(x)<\infty
\end{equation}
for some $a>0$. Then for any $p\geq 1$ the set ${\cal P}_n$ of
real polynomials in $n$ variables
is dense in $L^p( \mathbb R^n, \mu)$.
\end{theorem}

\begin{proof}
Since  $L^q(\mathbb R^n,\mu)$ for $\frac 1p+\frac 1q=1$ is the dual of $L^p(\mathbb R^n,\mu)$, in view of the Hahn-Banach
theorem it suffices to show that if $f\in L^q(\mathbb R^n,\mu)$ is such that
\begin{equation}\label{E2}
\int_{\mathbb R^n} x^k f(x)\,d\mu(x)=0
\end{equation}
for every multi-index $k\in \mathbb N^n_0$, then $f(x)=0$ a.e.
(Note that by (\ref{E1}) and the H\"older inequality, the integral
in (\ref{E2}) exists.) We denote
$$
A= \{(z_1,\ldots,z_n)\in\mathbb C^n\,|\, |\,{\rm Im}\, z_j|<\frac a
{2\sqrt{n}p}\ {\rm for\  all\ } j=1,\ldots,n\}.
$$
If $(z_1,\ldots,z_n)\in A$, using the Schwarz inequality we get
\begin{eqnarray*}
|\exp(-i\sum_{j=1}^nz_jx_j)|^p [\exp(\frac a{2p}\parallel x\parallel)]^p
&=& [ \exp(\sum_{j=1}^n\,{\rm Im}\,z_jx_j)\exp(\frac a{2p}\parallel x\parallel)]^p \\
&\leq &
 [ \exp(\sqrt{n} \frac a{2\sqrt{n}p}\parallel x\parallel)\exp(\frac a{2p}\parallel x\parallel)]^p
=e^{a\parallel x\parallel},
\end{eqnarray*}
and so by the H\"older inequality
\begin{equation}\label{E3}
\int_{\mathbb R^n} |\exp(-i\sum_{j=1}^nz_jx_j) f(x)| \,e^{\frac a{2p}\parallel x\parallel}\,d\mu(x) <\infty.
\end{equation}
We may thus define $F:A\to\mathbb C$ by the formula
\begin{equation}\label{E4}
F(z_1,\ldots,z_2) =
\int_{\mathbb R^n} \exp(-i\sum_{j=1}^nz_jx_j) f(x)\,d\mu(x),
\end{equation}
and using (\ref{E3}) we also find by induction that
\begin{equation}\label{E5} 
\frac{\partial^k}{\partial z_{j_1}\cdots\partial z_{j_k}} \, F(z_1,\ldots,z_n)
=\int_{\mathbb R^n} (-i)^k x_{j_1}\cdots x_{j_k}\, \exp(-i\sum_{j=1}^nz_jx_j) f(x)\,d\mu(x)
\end{equation}
for all $(z_1,\ldots,z_n)\in A$. 
(Differentiation under the integral sign is allowed in view of a standard argument
\cite[p.282]{Lang} based on the dominated convergence theorem and a general
mean value theorem \cite[p.103]{Lang}).)
Since $F$ is in each variable separately complex differentiable, it is in $A$ a holomorphic
function of $n$ complex variables.
(This is so by Hartogs's theorem but the more elementary Osgood  lemma in
\cite[p. 2]{GR} suffices here, as it is easy to show using the dominated convergence theorem
that  $F$ is continuous.)
Since each $\frac{\partial^k}{\partial z_{j_1}\cdots\partial z_{j_k}} \, 
F(0,\ldots,0)=0$
by assumption, the coefficients of the power series expansion of $F$ about the origin
vanish \cite[p. 3]{GR}, and so by the identity theorem \cite[p. 6]{GR},
$F$ is identically zero in $A$. In particular, for the Fourier--Stieltjes transform of the
complex measure $f\cdot \mu$ we have
$$
\int_{\mathbb R^n} \exp(-i\sum_{j=1}^nu_jx_j) f(x)\,d\mu(x)=0
$$
for all $(u_1,\ldots,u_n)\in\mathbb R^n$, and so $f\cdot \mu =0$, that is,
$f$ vanishes $\mu-a.e.$
\end{proof}

\begin{corollary}\label{corollary}
An exponentially bounded  measure $\mu:\cal B(\mathbb R^n):\to[0,\infty)$
 %satisfying the assumption {\rm (\ref{E1})}
is ultradeterminate in the sense of
\cite{Fuglede} and thus determinate.
\end{corollary}
\begin{proof}
See \cite[pp. 61, 58]{Fuglede}
\end{proof}

\section{Marginal measures and the uniqueness question}

In view of the importance of the final conclusion of Corollary 2.2
for our applications,
we develop in this section an alternative proof,  which does not
depend on the theory of holomorphic functions of several complex
variables. Some steps on the way have independent interest.

\begin{lemma}\label{le:3.1} Let $\mu$ be a Borel measure which
is determinate on $K \subseteq \mathbb R,$ where $K$
is one of the following possibilities:
\begin{enumerate}

\item[{\rm (i)}] $K = [a,\infty),$  $a \in \mathbb R$;

\item[{\rm (ii)}] $K =(-\infty,b],$ $b \in \mathbb R$;

\item[{\rm (iii)}] $K$ is a compact subset in $\mathbb R$.

\end{enumerate}
Then  
%$\mu^{K}$ is an N-extremal solution of the Hamburger
%problem on $\mathbb R$. In addition, 
the set of all polynomials
${\cal P}_1$ is dense in $L^2(K,\mu).$
\end{lemma}

{\bf Proof.} (i) 
If  $K = [0,\infty) $, then 
by      %the Riesz theorem \cite[Theorem ?]{Akh}
\cite[Corollary 3.9]{BeTh}, 
%$\mu^{[0,\infty)}$ is an N-extremal solution, and so
${\cal P}_1$ is dense in $L^2([0,\infty), \mu).$
Let  now $K=[a,\infty)$, and 
let $T_a:\ [0,\infty) \to[a,\infty)$ be the
translation defined by $T(u):= u+a.$
Put $\mu_a(X):= \mu \circ T_a(X)$ for any
Borel subset $X$ of $[0,\infty)$. 
Then $\mu_a$ is a Borel measure on $[0,\infty)$. 
Let $c_k := \int_a^\infty t^k d\mu(t)$ 
and $\tilde c_k := \int_0^\infty u^k d\mu_a(u)$. 
Then $c_k = \sum_{i=0}^k {k\choose i}a^{k-i}\tilde c_i$ 
and $\tilde c_k := \sum_{i=0}^k {k \choose i}
(-1)^{n-i}a^{k-i} c_i.$ 
Therefore, $\mu$ is determinate on $[a,\infty)$ 
if and only if $\mu_a$ is determinate on $[0,\infty)$.

(ii) This case can be proved by modifying the previous argument.

(iii) Let $K$ be a compact set in $\mathbb R.$ Then in
view of the Weierstrass approximation theorem, ${\cal P}_1$ is dense in $L^2(K, \mu)$. 
% so that ${\cal P}_1$ is dense in ${\cal
%L}_2(\mathbb R, \mu^{K})$.
%, and $\mu^{K}$ is an N-extremal solution.
\hfill{$\Box$}

\vspace{3mm}
The following result generalizes \cite{Petersen}, Theorem 3.

\begin{theorem}\label{th:3.2} Let $K = K_1 \times \cdots\times K_n$, where
each $K_i$ is either $\mathbb R$ or a
nonempty  subset of $\mathbb R$ satisfying
one of the conditions {\rm (i)--(iii)} in
Lemma {\rm \ref{le:3.1}}. Let
$\pi_i :\ K \to K_i$ be the $i$-th projection of $K$ onto
$K_i$, $i=1,\ldots,n.$ Then a measure $\mu\in  {\cal M}^*_n(K)$ is
determinate whenever all the projection measures
$\mu \circ \pi_i^{-1}$ are determinate on $K_i$ for $i=1,\ldots,n$.
\end{theorem}

{\bf Proof.} We only sketch the proof, the omitted details being essentially
the same as in the proof of Theorem 3 in \cite{Petersen}.
Let $\nu\in V[K,\mu]$.  Since by assumption the measures
$X\mapsto \mu(\pi_i^{-1}(X))$ are determinate, we get
$\nu \circ \pi_i^{-1} = \mu \circ \pi_i^{-1}.$ For any closed set
$K\subset\mathbb R$, let $C_c(K)$ denote the space of continuous real
 functions on $K$ with compact support.
For any real functions $f_i$ on $K_i$,  $i=1,\ldots, n,$ 
let  $f_1 \otimes \cdots \otimes f_n$ be defined by
$f_1 \otimes \cdots \otimes f_n(x_1,\ldots,x_n):= f_1(x_1)\cdots f_n(x_n).$
Using the density results of Lemma \ref{le:3.1}
we may show, following the argument of \cite{Petersen} referred to above,  
that, given $\epsilon >0$,  for each    $f_i \in C_c(K_i), i=1,\cdots,n $ 
we can find $n$ polynomials $p_i \in{\cal P}_1$, $i=1,\ldots,n,
$ such that
$$
\int_K |f_1\otimes \cdots
\otimes f_n - p_1\otimes\cdots\otimes p_n| d\nu < \epsilon. \eqno(2.3)
$$
For $f :\ K \to \mathbb R$,
define the extension $f^K$ of $f$ to
$\mathbb R^n$ via $f^K(x) = f(x)$ if $x \in K$, otherwise $f^K(x) = 0$. Let
$\mu^K$ be the extension of $\mu$ to ${\cal B}(\mathbb R^n)$ given by
$\mu^K(E) :=\mu(E \cap K), E\in{\cal B}\mathbb (R^n).$
Then $f \in L^1(K,\mu)$ if and only if
$f^K \in  L^1(\mathbb R^n,\mu^K)$,
and if this is  the case, then
$\int_K f d\mu = \int_{\mathbb R^n}f d\mu^K.$
It is known that the set
$\{f_1\otimes\cdots\otimes f_n:\, f_i \in C_c(\mathbb R)\}$ is dense
in $ L^1(\mathbb R^n,\mu^K)$.
For  $f \in C_c(\mathbb R),$ the restriction $f|_{K_i}$
of $f$ to $K_i$ gives a function in $C_c(K_i)$ for any
$i=1,\ldots,n$. 
Let  $f \in L^1(K,\nu)$, so that $f^K\in L^1(\mathbb R^n,\nu^K)$. 
For a given $\eta>0$
we can find $f_1,\ldots, f_n \in C_c(\mathbb R)$ such that
$\int_{\mathbb R^n} |f^K - f_1\otimes \cdots\otimes f_n| d\nu^K < \eta$.
Then $\int_K|f - f_1\otimes \cdots\otimes f_n\chi_K| d\nu =
 \int_{\mathbb R^n} |f^K - f_1\otimes \cdots\otimes f_n| d\nu^K < \eta$,
 which proves that the set $\{ f_1\chi_{K_1}\otimes\cdots  \otimes f_n\chi_{K_n}:\ f_i \in
C_c(\mathbb R)\} =\{f_1\otimes\cdots\otimes f_n: f_i \in C_c(K_i), i=1,\ldots,n\}$ is dense in $L^1(K,\nu).$
From (2.3) it easily follows
that ${\cal P}_n$ is dense in $L^1(K,\nu)$,
and therefore by a theorem of Douglas  \cite{Dou}, $\nu$ is an
extremal point of the convex set $V[K,\mu]$.
Since this is true for any $\nu\in V[K,\mu]$,
the set $V[K,\mu]$ has to be a singleton.
\hfill{$\Box$}

\begin{corollary}\label{co:3.4} Let $\mu $ be an exponentially bounded
  Borel measure on $K = K_1 \times \cdots\times K_n$, 
with $K_i$ as in Theorem {\rm \ref{th:3.2}},  that is,  
\begin{equation}\label{eq:2}
\int_{K} e^{a||x||}d\mu(x) <  \infty,
\end{equation}
for some    $a >0$.
Then all the moments
$c_k(\mu),$ $k \in \mathbb N_0^n,$
exist and are finite,
and the measure $\mu$ is determinate.
\end{corollary}

{\bf Proof.} 
By assumption, it is clear that
all the multidimensional moments $\int_K x^k d\mu(x)$
exist and are finite.  Fix $i=1,\ldots,n.$ Then
\begin{eqnarray*}
\int_{\mathbb K_i} e^{a|t|} d \mu \circ \pi_i^{-1}(t)  &=& \int_{K} e^{a|x_i|} d\mu(x_1,\ldots, x_n)\\
&\le& \int_{K} e^{a||x||} d\mu(x_1,\ldots,x_n ) < \infty.
\end{eqnarray*}
Using \cite[Theorem II.5.2]{Freud},  or \cite[Theorem 6]{BergChr},
we see that  $\mu \circ \pi_i^{-1}$ is determinate
for $i=1,\ldots,n.$  By Theorem \ref{th:3.2},
$\mu$ is determinate.
\hfill{$\Box$}

\section{Phase space observables}\label{S3}

Consider a phase space observable $A^T$,
defined by a state operator $T$, a positive trace one operator,
by means of the weakly defined integral
$$
%{\cal B}({\mathbb C})\ni Z\mapsto
A^{T}(Z):=\frac 1\pi\int_Z D_zTD_z^*\,d\lambda(z), \ Z\in{\cal B}
({\mathbb C}),
% \in\lh
$$
where $\lambda$ is the Lebesgue measure on the complex plane
$\mathbb C$, and
$D_z=e^{za^*-\overline{z}a}$, $z\in\mathbb C$,
is the unitary shift operator associated with the ladder operators
$a = \sum_{n\geq 0}\sqrt{n+1}\kb n{n+1}$ and
$a^* = \sum_{n\geq 0}\sqrt{n+1}\kb{n+1}{n}$ of  an  orthonormal basis
$(\ket n)_{n=0}^\infty$, called the number basis,
  of  a complex separable Hilbert space $\hi$.
Let $A^T_{\psi,\fii}$ denote the complex measure
$Z\mapsto\ip{\psi}{A^T(Z)\fii}$ defined by the
(positive normalized) operator measure $A^T$ and
the vectors $\psi,\fii\in\hi$, and let $\lh$ be the set of
bounded operators on $\hi$.

The {\em moment operators} of the operator measure $A^T$ are the linear operators
$$
A^T[m,n] :=\int_{\mathbb C}\,z^m\overline{z}^n\,dA^T(z),
$$
each defined on the linear subspaces
$$ %\begin{eqnarray*}
\mathcal D[m,n] =\{\fii\in\hi\,|\, z\mapsto z^m\overline{z}^n
\ {\rm is\ integrable\ w.r.t.\ } A^T_{\psi,\fii}\
{\rm for\ all }\ \psi\in\hi \},
$$%\end{eqnarray*}
and satisfying, for any $\fii\in\mathcal D[m,n], \psi\in\hi$,
$$
\ip{\psi}{A^T[m,n]\fii} = \int_{\mathbb C}\,z^m\overline{z}^n\,dA^T_{\psi,\fii}(z).
$$

We say that
the operator measure $A^T$ is  {\em determinate} if it is uniquely determined by its moment operators 
$A^T[m,n], m,n\geq 0$.

In a previous article \cite{DLY} we have investigated the moment problem for the polar coordinate
($\mathbb C\ni z= |z|e^{i\theta}, |z|\in[0,\infty),\theta\in[0,2\pi)$)
marginal measures of the phase space observables $A^{\ket s}$
associated with the number states $\ket s$, $s\in\mathbb N_0$. The operator
measures
\begin{eqnarray*}
&&\mathcal B([0,\infty))\ni R\mapsto A^{\ket s}(R\times[0,2\pi))\in\lh,\\
&&\mathcal B([0,2\pi))\ni X\mapsto A^{\ket s}([0,\infty)\times X)\in\lh
\end{eqnarray*}
were shown to be  determinate. Here we investigate the moment problem for the phase space observables
$A^{\ket s}$. %as well as for their Cartesian margins.

\begin{remark}\rm
The  complex moment problem of the measures
$\asff:\cal B(\mathbb C)\to[0,1]$ is here
interpreted as the $\mathbb R^2$-moment problem of $\asff:\cal B(\mathbb R^2)\to[0,1]$
via the identification $z=x+iy$. 
In \cite[Appendix]{StoSza98} the one-to-one correspondence of
the complex and the two-dimensional
moment sequences 
$\int_{\mathbb C}z^m\overline{z}^n\,\asff(d\lambda(z))$
and 
$\int_{\mathbb R^2}x^my^n\,\asff(dxdy)$, with $z=x+iy$,
has been demonstrated.
\end{remark}

\section{On the moment operators of $A^{\ket s}$}

To determine
the moment operators $A^{\ket s}[m,n], m,n\geq 0$, of a phase space
observable $A^{\ket s}$
defined by a number state $\ket s$, we first observe that
for any $m,n$,
and for any number states $\ket k, \ket l$, the integral
$$
\ip{k}{\asmn|l} =\int_{\mathbb C}\,z^m
\overline{z}^n\,dA^{\ket s}_{\ket k,\ket l}(z)
$$
exists and is finite.
Indeed, by a direct computation one gets
\begin{eqnarray}
&&\int_{\mathbb C}\,|z|^{m+n}\,dA^{\ket s}_{\ket k,\ket l}(z)
=
\frac 1\pi \int_{\mathbb C}\,|z|^{m+n}\,\ip{k}{D_z|s}\ip{s}{D_z^*|l}\,d
\lambda(z) =\label{Eint}\\
&&
\frac {\delta_{k,l}}{s!}
\sum_{r=0}^{[k,s]}\sum_{r'=0}^{[k,s]}
%(-1)^{r+r'}\binom sr\binom s{r'}\frac{k!}{(k-r)!(k-r')!}
a(s,k,r)a(s,k,r')
\int_0^\infty e^{-|z|^2}|z|^{m+n+2(s+k-r-r')}\,2|z|d|z|<\infty,\nonumber
% \\  &&{}\qquad\qquad\qquad\qquad \qquad\qquad\qquad\qquad \qquad\qquad
\end{eqnarray}
where $$a(s,k,r)= (-1)^{s-r}\binom sr \sqrt{k!}/(k-r)!,$$ and
where  $[k,s]$ denotes the minumum of $k$ and $s$ \cite{LMY}.

We recall from \cite[Lemma A.2]{LMY} that by the positivity of the
operator measure $A^T$,
that is, by the fact that any $A^T(Z), Z\in\cal B(\mathbb C)$, is a
positive operator,
the domain
$\mathcal D[m,n]$ of $A^T[m,n]$ contains as a subspace the set
$$
\widetilde{\mathcal D}[m,n] =\{\fii\in\hi\,|\, z\mapsto
|z|^{2(m+n)}\ {\rm is\ integrable\ w.r.t.\ } A^T_{\fii,\fii} \}.
$$
Since $A^T$ is {\em not} projection valued, the set
$\widetilde{\mathcal D}[m,n]$ could be a proper
subset of $\mathcal D[m,n]$.

The above result (\ref{Eint})  shows  that for any $A^{\ket s}$
$$
{\rm lin}\,\{\ket k\,|\,k\in\mathbb N_0\} \subset
\widetilde{\mathcal D}[m,n]
\subset {\mathcal D}[m,n],
$$
showing that all the moment operators $\asmn$ are densely defined.
Denoting
$$
\overline{\asmn} :=\int_{\mathbb C} \overline{z^m\overline{z}^n}\,
dA^{\ket s}(z)
$$
we observe that
$\overline{\asmn} %\int_{\mathbb C} \overline{z}^mz^n\,dA^{\ket s}(s)
=\asnm$, as well as
$\overline{\asnm} =\asmn$. Therefore, using \cite[Lemma A4]{LMY}, we see
that
the adjoint  of $\asmn$, resp. $\asnm$,  is an extension of
$\overline{\asmn}$, resp. $\overline{\asnm}$, that is,
\begin{eqnarray}
&&\asmn\subseteq \asnm^*, \label{exten1}\\
&&\asnm\subseteq \asmn^*.\label{exten2}
\end{eqnarray}

The matrix elements of the operators $\asmn$ in the
number basis can easily be computed,
and we get \cite{LMY}:
$$
\ip{k}{\asmn|l} =
\frac 1\pi\frac 1{s!}\sum_{r=0}^{[k,s]}\sum_{r'=0}^{[l,s]}
a(s,k,r)a(s,l,r')\, I(m,n,s,k,l,r,r')
$$
where
\begin{eqnarray*}
I(m,n,s,k,l,r,r')&=& \int_{\mathbb C} e^{-|z|^2}\, z^{m+k+s-r-r'}\overline{z}^{n+l+s-r-r'}\,d\lambda(z)\\
&=& 0,\  {\rm\ whenever\  }\  k+m\ne l+n, \\
&=& \pi\,(m+s+k-r-r')!, \ {\rm\ for\  }\  k+m= l+n.
\end{eqnarray*}
Therefore,
\begin{eqnarray*}
\ip{k}{\asmn|l} &=&  0, \  {\rm\ for\  }\  k+m\ne l+n, \\
%&&\ip{k}{\asmn|m-n+k} =\\
%&&\frac{k!(m-n+k)!}{s!}
&=&
\frac 1{s!} \sum_{r=0}^{[k,s]}\sum_{r'=0}^{[m-n+k,s]}
a(s,k,r)a(s,m-n+k,r')(m+s+k-r-r')!, \\
&&\ {\rm\ for\  }\  k+m= l+n.
%(-1)^{r+r'} \binom sr\binom s{r'}
%\frac{(m+s+k-r-r')!}{(k-r)!(m-n+k-r')!},
\end{eqnarray*}
It seems difficult to determine the explicit form of the operators $\asmn$.
However, it is known  \cite{LMY} that
\begin{eqnarray*}
&&\asnn = \sum_{i,j=0}^n a_{ij} s^{n-j} N^i,\  a_{ij}\   {\rm integers},
\   \mathcal D(\asnn) =\mathcal D(N^n),\   N=a^*a, \\
%&&\qquad\qquad\qquad\qquad a_{ij}\  \ {\rm integers}\\
&&A^s[n,0] = a^n,\  A^s[0,n]= (a^*)^n,\  \   \cal D(A^s[n,0])=\cal D(A^s[0,n]) = \cal D(a^n).
\end{eqnarray*}
For  $\ket s =\ket 0$ one may quickly confirm that
$$
\ip{k}{\agmn|l} = \ip{k}{a^m(a^*)^n|l}
$$
for any $m,n\in\mathbb N_0$, and for any number states $\ket k,\ket l$.
Moreover,
one easily  shows that $\cal D(a^m(a^*)^n)\subseteq  \widetilde{\mathcal D}[m,n]$
and that, actually, $\agmn$ extends the operator $a^m(a^*)^n$, which, together with the
above relations (\ref{exten1}--\ref{exten2})  shows that
$$
\agmn = a^m(a^*)^n.
$$
The possibility of obtaining the operators $a^m(a^*)^n$
from the "diagonal coherent state representation" $\frac 1\pi\int_{\mathbb C} z^m\overline{z}^n\kb zz\,d\lambda(z)$
was perhaps first noticed by Sudarshan \cite{Sudarshan}.
The papers \cite{Cahill-Glauber, Agarwal-Wolf} are further
elaborations  on the related  `phase space
quantization methods'.  From the point of view of the theory of
operator integrals these pioneering papers amounted to showing that  $a^m(a^*)^n\subset\agmn$.

\section{The uniqueness of $A^{\ket s}$}
\label{uniqueness}

We show next that the phase space observable $A^{\ket s}$ is
uniquely determined by its moment operators $\asmn$, $m,n\in\mathbb N_0$.
In other words,  if $E:\cal B(\mathbb C)\to\lh$ is another
normalised positive operator
measure such that its moment operators equal those of $A^{\ket s}$,
that is,
$E[m,n]=\asmn$ for all $m,n\in\mathbb N_0$, then $E=A^{\ket s}$.
Actually,  the equality $E=A^{\ket s}$
already follows if the moment operators of $E$ agree with those of $A^{\ket s}$ on a dense subset.

Let $\ket k,\ket l$ be any two number states and consider the complex measure
$A^{\ket s}_{\ket k,\ket l}$. Its values are
\begin{eqnarray*}
&&A^{\ket s}_{\ket k,\ket l}(Z) =\frac 1\pi\int_Z\ip{k}{D_z|s}\ip{s}{D_z^*|l}\,d\lambda(z) =\\
&& \frac 1\pi\frac 1{s!}\sum_{r=0}^{[k,s]}\sum_{r'=0}^{[l,s]}
a(s,k,r)a(s,l,r')\int_Z e^{-|z|^2}z^{s+k-r-r'}\overline{z}^{s+l-r-r'}\,d\lambda(z).
\end{eqnarray*}
%where $a(s,k,r)= (-1)^{s-r}\binom sr \sqrt{k!}/(k-r)!$.
In particular, for each $\ket k$ the probability measure
$A^{\ket s}_{\ket k,\ket k}$ has the density
$$
\frac 1\pi\frac 1{s!}\sum_{r=0}^{[k,s]}\sum_{r'=0}^{[k,s]}
a(s,k,r)a(s,k,r')e^{-|z|^2}|z|^{2(s+k-r-r')}
$$
with respect to the Lebesgue measure $\lambda$.
But then for any $a\in\mathbb R$,
\begin{eqnarray*}
\int_{\mathbb C} e^{a|z|}\,dA^{\ket s}_{\ket k,\ket k}
&=&
\frac 1\pi\frac 1{s!}\sum_{r=0}^{[k,s]}\sum_{r'=0}^{[k,s]}
a(s,k,r)a(s,k,r')
\int_{\mathbb C} e^{a|z|}e^{-|z|^2}|z|^{2(s+k-r-r')}\,d\lambda(z)\\
&=&
\frac 1{s!}\sum_{r=0}^{[k,s]}\sum_{r'=0}^{[k,s]}
a(s,k,r)a(s,k,r') \int_{0}^\infty
e^{a|z|}e^{-|z|^2}|z|^{2(s+k-r-r')}\,2|z|d|z|\\
&=&
\frac 1{s!}\sum_{r=0}^{[k,s]}\sum_{r'=0}^{[k,s]}
a(s,k,r)a(s,k,r') e^{(a/2)^2}
\int_{0}^\infty e^{-(|z|-a/2)^2} |z|^{2(s+k-r-r')}\,2|z|d|z|\\
&<&\infty.
\end{eqnarray*}
By Corollary~\ref{corollary} each  $A^{\ket s}_{\ket k,\ket k}$
is determinate, that is, $|V(\mathbb C,A^{\ket s}_{\ket k,\ket k})|=1$.

For any $\ket k,\ket l$, $k\ne l$, $c\in\mathbb C, |c|=1$, we also have
$$%\begin{eqnarray*}
%\int_{\mathbb C} e^{a|z|}\,dA^{\ket s}_{\ket k,\ket l}
%&=& \frac 14\sum_{r=0}^3i^r
\int_{\mathbb C} e^{a|z|}\,dA^{\ket s}_{\ket l+c\ket k,\ket l+c\ket k}=
%\frac 14\sum_{r=0}^3i^r   \left(
\int_{\mathbb C} e^{a|z|}\,dA^{\ket s}_{\ket l,\ket l}
+
\int_{\mathbb C} e^{a|z|}\,dA^{\ket s}_{\ket k,\ket k}%\right)
<\infty,
$$%\end{eqnarray*}
since for instance
$$
\int_{\mathbb C} e^{a|z|}\,dA^{\ket s}_{\ket k,\ket l}=0.
$$
Thus
all the measures $A^{\ket s}_{\ket l+c\ket k,\ket l+c\ket k}$ are determinate.

Assume now that  $E:\mathcal B(\mathbb C)\to\lh$ is another operator measure
for which
 $E[m,n]=A^{\ket s}[m,n]$ on ${\rm lin}\,\{\ket k\,|\, k\in\mathbb N_0\}$.
Using the  polarization identity we  get
for all number states $\ket k$ and $\ket l$,
\begin{eqnarray*}
E_{\ket k,\ket l}&=&
\frac 14\sum_{r=0}^3i^rE_{\ket l+i^r\ket k,\ket l+i^r\ket k}\\
&=&
%\sum_{k,l}\ip{\psi}{k}\ip{l}{\fii}\ip{k}{A^{\ket s}(Z)|l} =\ip{\psi}{A^{\ket s}(Z)\fii}
\frac 14\sum_{r=0}^3i^rA^{\ket s}_{\ket l+i^r\ket k,\ket l+i^r\ket k}
=A^{\ket s}_{\ket k,\ket l}.
\end{eqnarray*}
This shows that $E=A^{\ket s}$, that is,  the phase space observable
$A^{\ket s}$ defined by the number state $\ket s, s\in\mathbb N_0$,  is determinate.

\section{The uniqueness of $A^{\ket s}$ through its Cartesian margins}
\label{margins}

Using Theorem 3 of Petersen \cite{Petersen},
the  uniqueness $A^{\ket s}$
may also be obtained from the determinacy of its Cartesian marginal measures.
We shall demonstrate this result next.

To facilitate the calculations,
we pass to the $L^2(\mathbb R)$-realization of the phase space observables $A^{\ket s}$.
Let $W:\hi\to L^2(\mathbb R)$ be the unitary mapping for which $W(\ket n)=f_n$, $n\in\mathbb N_0$, where
$f_n$ is the $n$-th Hermite function,
\begin{eqnarray*}
f_n(x)&=& N_n\,e^{-x^2/2}\,H_n(x),\  x\in\mathbb R,\\
N_n&=& (\sqrt{\pi}\, 2^nn!)^{-1/2},\\
H_n(X)&=&(-1)^ne^{x^2}\frac{d^n}{dx^n}\,e^{-x^2},\ x\in\mathbb R.
\end{eqnarray*}
When we identify
$\mathbb C$ with $\mathbb R^2$, and write
$z=\frac{q+ip}{\sqrt 2}$, the phase space observable
$A^{\ket s}$, defined by $f_s$, gets the form
$$
A^{\ket s}(Z) = \frac 1{2\pi}\int_Z\kb{e^{-iqP+ipQ}f_s}{e^{-iqP+ipQ}f_s}\,dqdp,
$$
with $(Q,P)$ being the Schr\"odinger pair on $L^2(\mathbb R)$ \cite{GW}.
%(Pit"isk"h"n selitt"" tarkemmin??)
The Cartesian marginal measures of $A^{\ket s}$
are known to be the unsharp position $\uqs$ and  the unsharp momentum $\ups$, with
\begin{eqnarray*}
\uqs(X) &=& (\chi_X*|f_s|^2)(Q), \ X\in\mathcal B(\mathbb R),\\
\ups(Y) &=& (\chi_Y*|\hat{f_s}|^2)(P), \ Y\in\mathcal B(\mathbb R),
\end{eqnarray*}
respectively,
where $\chi_X*|f_s|^2$ is the convolution of the characteristic function $\chi_X$ with the
density
function $|f_s|^2$, and $\hat{f_s}$ is the Fourier transform of $f_s$, see, e.g. \cite[Theorem 3.4.1]{Davies}.

Let $\fii\in L^2(\mathbb R)$ be a unit vector, and consider the probability measure $\asff$. Its Cartesian
marginal probability measures are $\uqsf$ and $\upsf$, respectively.
Clearly, they are absolutely continuous with respect to the Lebesgue measure of $\mathbb R$.
Let $\gqsf$ be the Radon-Nikodym derivative of $\uqsf$ with respect to $dq$.
We assume now that $\fii\in\cal C^{\infty}_0(\R)$ so that we may take
$$
\gqsf(x)=\int_\R |f_s(x-q)|^2|\fii(q)|^2\,dq.
$$
Let ${\rm supp}\,g\subseteq[a,b]$, $M\in[0,\infty)$, be such that $|\fii(q)|^2\leq M$ for all $x\in[a,b]$, and let
$|q|\leq C$, $a\leq q\leq b$. Then
\begin{eqnarray*}
\gqsf(x)&=& N_s^2\int_a^b e^{-(x-q)^2}H_s(x-q)^2|\fii(q)|^2\,dq\\
&\leq& M\,N_s^2 \, e^{-x^2}\int_a^b e^{-q^2+2qx}H_s(x-q)^2\,dq\\
&\leq& M\,N_s^2 \, e^{-x^2} e^{2C|x|}p_{2s}(x),
\end{eqnarray*}
where $p_{2s}$ is a polynomial of $x$ of degree $2s$. But then for any $a>0$,
$$
\int_\R e^{a|x|}e^{-x^2} e^{2C|x|}p_{2s}(x)\,dx < \infty,
$$
which shows that the probability measure $\uqsf$ is exponentially
bounded.  Therefore, all the moments of the
probability measure  $\uqsf$ are finite and the measure
$\uqsf$ is determinate for each unit vector $\fii\in\cal C^{\infty}_0(\R)$.
Similarly, any probability
measure $\upsf$, $\fii\in\cal C^{\infty}_0(\R)$,
$\parallel\fii\parallel=1$, is determinate,
so that, by \cite[Theorem 3]{Petersen}, or by Theorem \ref{th:3.2}, any phase space
probability measure $\asff$,
$\fii\in\cal C^{\infty}_0(\R)$, $\parallel\fii\parallel=1$, is determinate.

\begin{remark}\rm
Not all the probability
measures $\uqsf$, resp. $\upsf$,
$f\in L^2(\R), ||f|| =1$, can
be determinate since the moment operators
of $\uqs$ are unbounded operators.
Since the phase space observable $A^{\ket s}$ is known to be
informationally complete \cite{AliPru,BCL}
(that is, for any two state operators $T,U$,
if $A^{\ket s}_T=A^{\ket s}_U$, then $T=U$),
it also follows that if $\psi$ and $\fii$ are two
different vector states such that
$\uqs_{\psi,\psi}=\uqs_{\fii,\fii}$ and
$\ups_{\psi,\psi}=\ups_{\fii,\fii}$, then the
measures $\uqs_{\psi,\psi}$ and $\ups_{\psi,\psi}$
cannot both be determinate, since otherwise
also $\aspp$ (as well as $\asff$) would be determinate,
with the implication that the states $\kb{\psi}{\psi}$
and $\kb{\fii}{\fii}$ would be the same,
which need not be the case, see, e.g. \cite[Sect. 2.3]{BL}.
\end{remark}

Assume now that $E:\cal B(\mathbb C)\to \cal L(L^2(\R))$ is another positive operator measure such that
$E_{\fii,\fii} =\asff$ for all $\fii\in\cal C^{\infty}_0(\R)$, $\parallel\fii\parallel=1$.
Let $\psi$ be any unit vector of $L^2(\mathbb R)$. Since $\cal C^{\infty}_0(\R)$ is dense in $L^2(\mathbb R)$, $\psi$
 either is in $\cal C^{\infty}_0(\R)$ or  a limit of a sequence of vectors $\fii_n\in\cal C^{\infty}_0(\R)$.
Let $\psi=\lim\fii_n$.  
Then $\lim E_{\fii_n,\fii_n}(Z)=E_{\psi,\psi}(Z)$ 
as well as $\lim E_{\fii_n,\fii_n}(Z)=A^{\ket s}_{\psi,\psi}(Z)$
uniformly for
 $Z\in\cal B(\mathbb C)$, which implies that $E_{\fii_n,\fii_n}\to E_{\psi,\psi}$
and $E_{\fii_n,\fii_n}\to A^{\ket s}_{\psi,\psi}$
in the total variation norm \cite[p. 97]{DunSch}. Therefore, $E_{\psi,\psi}=A^{\ket s}_{\psi,\psi} $ for
any unit vector $\psi\in L^2(\mathbb R)$. By the
polarization identity, the operator measures $E$ and
$A^{\ket s}$ are the same. To conclude, we have established
the following result.

\begin{corollary}
Let $A^{\ket s}$ be the phase space observable defined by the number state $\ket s = W^{-1}f_s$, $s\in\mathbb N_0$.
The moment operators $A^{\ket s}[m,n]$ are densely defined,
$W^{-1}(C_0^\infty(\R))\subset\widetilde{\cal D}[m,n]\subseteq\cal D[m,n]$, and the observable
 $A^{\ket s}$ is uniquely determined by the restrictions of its moment operators to $W^{-1}(C_0^\infty(\R))$.
\end{corollary}

\section{The uniqueness of $A^{\ket s}$ in terms of its polar coordinate
margins}

In addition to the complex moments - which, as we have seen, essentially
amount to the real moments in terms of the Cartesian representation -
of a phase space observable, it is illuminating to consider the
real moments in terms of the polar coordinate representation. This we
do next making use of the generalization of Petersen's result
expounded in Theorem \ref{th:3.2}.

Consider the phase space observable $A^{\ket s}$ and its polar coordinate moment
operators $\int_0^\infty\int_0^{2\pi}r^n\theta^m\, dA^{\ket s}(re^{i\theta})$. The polar coordinate marginal measures
are
\begin{eqnarray*}
&&{\cal B}([0,\infty))\ni R\mapsto A^{\ket s}(R\times[0,2\pi))\in\lh,\\
&&{\cal B}([0,2\pi))\ni X\mapsto A^{\ket s}([0,\infty)\times X)\in\lh,
\end{eqnarray*}
the second of them being compactly supported and thus determinate. In 
\cite[Section 5]{DLY} it was shown that also the radial margin of $A^{\ket s}$ is uniquely determined by
its (unbounded self-adjoint) moment operators. Therefore, by Theorem \ref{th:3.2}, we may  conclude that
the phase space observable  $A^{\ket s}$ is uniquely determined also by its polar coordinate moment operators
$\int_0^\infty\int_0^{2\pi}r^n\theta^m\, dA^{\ket s}(re^{i\theta}), n,m\in\mathbb N_0$.
The same conclusion can also be obtained from \cite[Theorem 3.6]{BeTh} concerning rotation invariant
moment problem.
We do not pursue to determine the moment operators 
$\int_0^\infty\int_0^{2\pi}r^n\theta^m\, dA^{\ket s}(re^{i\theta})$, since their physical relevance is less direct.

\vspace{3mm}

{\bf Acknowledgement.} One of the authors (AD) is thankful  to the
Academy of Finland for organizing his stay at the University of
Turku, in October 2000. The paper was partially supported by the grant
VEGA 2/7193/00, Slovakia.

%%%%%%%%%%%%%%%%%%%%%%%%%%%


\begin{thebibliography}{References}{}

\bibitem{Agarwal-Wolf}
G.S. Agarwal, E. Wolf,
Calculus for functions of noncommuting operators and general
phase-space methods in quantum mechanics, I,
{\em Phys. Review D} {\bf 2} (1970), 2161--2186.

\bibitem{Akh}
N. Akhiezer,
{\it The Classical Moment Problem},
Edinburg, Oliver and Boyd, 1965.

\bibitem{AG1}
N.I. Akhiezer, I.M. Glazman,
{\em Theory of Linear Operators in Hilbert Space, Vol. 1},
Fredrik Ungar, New York, 1961.

\bibitem{AG2}
N.I. Akhiezer, I.M. Glazman,
{\em Theory of Linear Operators in Hilbert Space, Vol. 2},
Fredrik Ungar, New York, 1963.

\bibitem{AliPru}
S.T. Ali, E. Progovecki,
{\em Physics} {\bf 89A} (1977), 501--521.

\bibitem{BergChr}
C. Berg, J.P.R. Christensen,
Density questions in the classical theory of moments,
{\it Ann. Inst. Fourier, Grenoble} {\bf 31} (1981), 99--114.

\bibitem{BeTh} C. Berg, M. Thill, Rotation invariant
moment problems, {\em  Acta Math.} {\bf 167} (1991), 207--227.



\bibitem{BCL}
P. Busch, G. Cassinelli, P. Lahti,
Probability structures for quantum state spaces,
{\em Rev. Math. Phys.} {\bf 7} (1995), 1105--1121.

\bibitem{BL}
P. Busch,  P. Lahti,
The determination of the past and future of a physical system in quantum
mechanics, {\em Found. Phys.} {\bf 19} (1989), 633--678.

\bibitem{Cahill-Glauber}
K.E. Cahill, R.J. Glauber,
Ordered expansions in boson amplitude operators,
{\em Phys. Review} {\bf 177} (1969), 1857--1881.

\bibitem{Davies} E.B. Davies,
{\it The Quantum Theory of Open Systems}, Academic Press, 1976.

\bibitem{Dou} R.D. Douglas, On extremal measures and
subspace density, {\em  Mich. Math. J.} {\bf 11} (1964), 243--246.

\bibitem{DunSch}
N. Dunford, J.T. Schwartz,
{\it Linear Operators. Part II},
Interscience Publishers, John Wiley \& Sons, 1963.

\bibitem{DLY}
A. Dvure\v censij, P. Lahti, K. Ylinen,
Positive operator measures determined by their moment operators,
{\it Rep.  Math. Phys.} {\bf 45} (2000), 139--146.

\bibitem{Freud}
G. Freud,
{\it Orthogonal Polynomials}, Akad\'{e}miai Kiad\'{o}, Budapest, 1971.



\bibitem{Fuglede}
B. Fuglede,
The multidimensional moment problem,
{\em Expos. Math.} {\bf 1} (1983), 47--65.


\bibitem{GW}
J.C. Garrison, J. Wong,
Canonically conjugate pairs, uncertainty relations, and phase operators,
{\em J. Math. Phys.} {\bf 11} (1970) 2242-2249.

\bibitem{GR}
R.C. Gunning, H. Rossi,
{\em Analytic Functions of Several Complex Variable}, Prentice-Hall, 1965.

\bibitem{Hew}
E. Hewitt,
Remark on orthonormal sets in $\cal L_2(a,b)$,
{\it Amer. Math. Monthly} {\bf 61} (1954), 249--250.

\bibitem{LMY}
P. Lahti, M. M\c aczy\'nski, K. Ylinen,
The moment operators of phase space observables and their number margins,
{\it Rep. Math. Phys.} {\bf 41} (1998), 319--331.

\bibitem{Lang}
S. Lang,
{\em Analysis II}, Addison-Wesley Publ. Co., Inc., Reading, 1969.

\bibitem{Petersen}
L.C. Petersen,
On the relation between the multidimensional moment problem and
the one-dimensional
moment problem,
{\it Math. Scand.} {\bf 51} (1982), 361--366.

\bibitem{StoSza98}
J. Stochel, F.H. Szafraniec,
The complex moment problem and subnormality: a polar decomposition approach,
{\em J. Funct. Anal.} {\bf 159} (1998), 432--491.

\bibitem{Sudarshan}
E.C.G. Sudarshan,
Equivalence of semiclassical and quantum mechanical descriptions of
statistical light beams,
{\em Phys. Review Letters} {\bf 10} (1963), 277--279.

\end{thebibliography}
\end{document}